# Two Dimensional Atomic Crystals

K. S. Novoselov[1], D. Jiang[1], T. Booth[1], V.V. Khotkevich[1], S. M. Morozov[2], A. K. Geim[1]

[1]School of Physics and Astronomy, University of Manchester, Manchester, M13 9PL, UK
[2]Institute for Microelectronics Technology, 142432, Chernogolovka, Russia

*We report free-standing atomic crystals that are strictly 2D and can be viewed as individual atomic planes pulled out of bulk crystals or as unrolled single-wall nanotubes. By using micromechanical cleavage, we have prepared and studied a variety of 2D crystals, including single layers of boron nitride, graphite, several dichalcogenides and complex oxides. These atomically-thin sheets (essentially gigantic 2D molecules unprotected from the immediate environment) are stable under ambient conditions, exhibit high crystal quality and are continuous on a macroscopic scale.*

Dimensionality is one of the most defining materials parameters such that the same chemical compound can exhibit dramatically different properties, depending on whether it is arranged in a 0, 1, 2 or 3 dimensional (D) crystal structure. While quasi-0D (e.g., cage molecules [1]), quasi-1D (e.g, nanotubes [2-4]) and, of course, 3D crystalline objects are well documented, dimensionality 2 is conspicuously absent among experimentally known crystals. On the other hand, there are many layered materials with strong in-plane bonds and weak, van der Waals-like coupling between layers. Due to this layered structure, it has long been tempting to try splitting such materials into individual atomic layers, although it remained unclear whether free-standing atomic layers could exist in principle (thin films become thermodynamically unstable – decompose or segregate – below a certain thickness, typically, of many dozens layers). So far, most efforts have focused on chemical exfoliation of strongly-layered materials and, in particular, of stage-I intercalated graphite [5]. During exfoliation, monolayers must at some moment separate from each other. However, no 2D crystals have ever been isolated from the resulting slurries, possibly because single layers appear only as a transient state and involve detachments over microscopic regions. Indeed, latest studies of chemically-exfoliated graphite have shown that its sediments consist of restacked and scrolled multilayer sheets rather than of individual monolayers [6-8]. An alternative approach has been the use of mechanical cleavage [9-14]. Earlier reports described mechanically-cleaved flakes consisting of tens and hundreds of layers, but the recently renewed interest in thin graphitic films led to materials with thickness of just a few graphene layers [12-15]. Now we have developed a cleavage technique that extends the approach to its limit, allowing isolation and study of individual crystal planes from a variety of strongly-layered materials.

Figure 1 shows several examples of cleaved samples and illustrates that these are only one atomic layer thick but nearly macroscopic laterally. To extract such 2D crystallites, we used a simple but effective procedure. A fresh surface of a layered crystal was rubbed against another surface (virtually any solid surface is suitable), which left a variety of flakes attached to it (the rubbing process can be described as similar to "drawing by chalk on a blackboard"). Unexpectedly, amongst the resulting flakes we always found single layers. Their preliminary identification amid thicker flakes and other residue was done in an optical microscope. 2D crystallites become visible on top of an oxidized Si wafer (Fig. 1d), because even a monolayer adds up sufficiently to the optical path of reflected light so that the interference colour changes with respect to the one of an empty substrate (phase contrast). The whole procedure takes literally half an hour to implement and identify probable 2D crystallites. Their further analysis was done by atomic force microscopy (AFM), where single-layer crystals were selected as those exhibiting an apparent [12] thickness of about the interlayer distance in the corresponding 3D crystals.

Despite its simplicity, the described cleavage technique has several non-obvious features that are instructive to analyze, as this also allows one to understand why 2D crystals were not discovered earlier (see, e.g., refs [9-11,13,14] where mechanically-cleaved graphitic flakes 10 to 100 layers thick were reported). First, monolayers are in a great minority amongst accompanying thicker flakes. Second, unlike nanotubes, 2D crystals have no clear signatures in transmission electron microscopy [6-8]. Third, monolayers are completely transparent to visible light and cannot be seen in an optical microscope on most substrates (e.g., on glass or metals). Forth, AFM is currently the only methods allowing definitive identification of single-layer crystals but it has a very low throughput (especially, for the case of the high-resolution imaging required) and, in practice, it would be impossible to find cleaved 2D crystallites by scanning surfaces at random. Finally, as mentioned earlier, it was not obvious that isolated atomic planes could survive without their parent crystals (for example, mechanically-cleaved quasi-1D $NbSe_3$ crystallites ≈100 nm in diameter were found to deteriorate rapidly [16]). With the benefit of hindsight, the critical step that allowed us to find 2D crystallites is the discovered possibility of their tentative identification in an optical microscope.

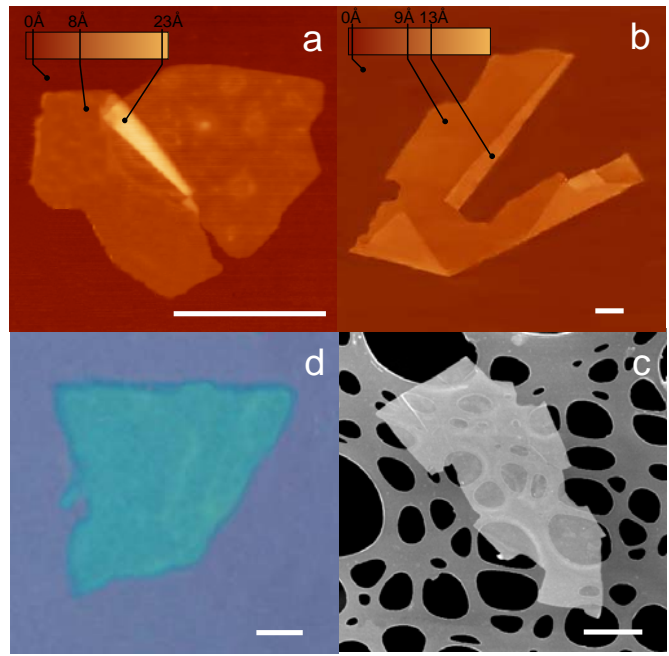

Figure 1. Two dimensional crystal matter. Single-layer crystallites of (**a**) $NbSe_2$, (**b**) graphite, (**c**) $Bi_2Sr_2CaCu_2O_x$ and (**d**) $MoS_2$ visualized by AFM (a,b), SEM (c) and in an optical microscope (d). All scale bars are 1μm. The 2D crystallites are on top of an oxidized Si wafer (300 nm of thermal $SiO_2$) (a-c) and on top of a holey carbon film (d). Note that 2D crystallites were often raised by an extra few Å above the supporting surface, probably due to a layer of absorbed water. In such cases, the pleated and folded regions seen on many AFM images and having the differential height matching the interlayer distance in the corresponding 3D crystals help to distinguish between single- and double-layer crystals.

Representative samples of several 2D materials (namely, of BN, $MoS_2$, $NbSe_2$, $Bi_2Sr_2CaCu_2O_x$ and graphite) obtained and identified by the procedures described above were further investigated by scanning tunnelling, scanning electron and high-resolution transmission electron microscopy (STM, SEM and HRTEM, respectively). Figure 2 shows examples of the obtained atomic-resolution images. These studies [17] confirmed that the prepared 2D crystallites remained monocrystalline under ambient conditions and no degradation was noticed over periods of many weeks. Within experimental resolution, the crystal structure of isolated layers remained the same as for stacked layers within 3D crystals. Note that 2D $Bi_2Sr_2CaCu_2O_x$ showed a superstructure with a unidirectional modulation period of ≈28Å, which is similar to the superstructure observed in thinned samples of bulk $Bi_2Sr_2CaCu_2O_x$ prepared for HRTEM [18].

We have also investigated electrical conductivity of the selected five 2D materials to assess their microscopic quality and macroscopic continuity. This was done by using field-effect-transistor-like devices such as the one shown in the inset of Fig. 3 (devices were prepared by electron-beam lithography). 2D Bi$_2$Sr$_2$CaCu$_2$O$_x$ and BN were found to be highly insulating, and no induced conductivity was detected even at gate voltages as high as 0.3V/nm, i.e. close to the electrical breakdown of SiO$_2$. This probably indicates that band gaps in these 2D materials are larger than in SiO$_2$. We also tried annealing single-layer Bi$_2$Sr$_2$CaCu$_2$O$_x$ in oxygen but the crystals always remained insulating.

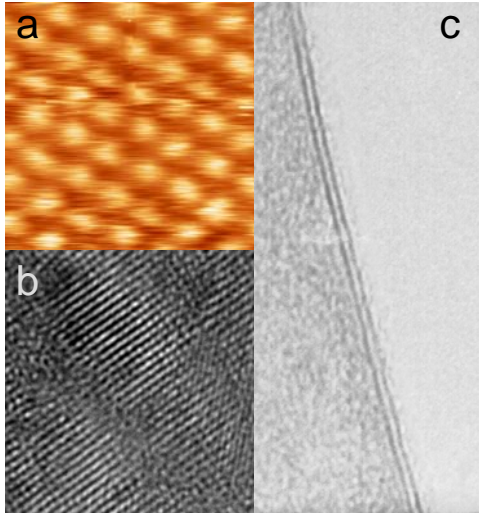

Figure 2. Atomic-resolution images of 2D materials. (**a**) Unfiltered STM image of the crystal lattice in NbSe$_2$ monolayer on top of an oxidized Si wafer. Note that for the STM measurements an Au film was deposited around 2D crystallites to provide an electrical contact. (**b**) HRTEM images of the 2D Bi$_2$Sr$_2$CaCu$_2$O$_x$ crystal shown in Fig. 1c. (**c**) HRTEM image of a double-layer MoS$_2$. This image is shown to make a connection between our approach based on AFM identification of 2D crystals and the traditional HRTEM approach used for quasi-1D materials (all nanotubes were first found by using HRTEM, where dark lines indicating nanotube's walls parallel to the electron beam are easily visible). No similar signature exists for 2D crystals (see refs. [6-8]), and we also found it difficult to align 2D samples exactly parallel to the electron beam. However, for two-layer crystals, their thickness is easily identifiable not only in AFM but also in HRTEM due to folded regions seen as two dark lines (in the case of Fig. 2c, the separation is ≈6.5Å, in agreement with the interlayer distance in bulk MoS$_2$). Occasionally, we did observe short dark lines (c.f., [8]) that might be folded monolayers but no independent proof for this (by simultaneous AFM studies) has been obtained so far.

On the contrary, 2D graphite (graphene) and both 2D dichalcogenides were found metallic and exhibited a pronounced electric field effect (Fig. 3). Their carrier mobilities were determined as $\mu = \sigma(V_g)/en(V_g)$ where $e$ is the electron charge and $n \propto V_g$ is the carrier concentration induced by gate voltage $V_g$ ($n \approx 7.2 \times 10^{10}$ cm$^{-2}$/V for 300 nm SiO$_2$). As seen from Fig. 3, $\sigma$ was proportional to $V_g$ over large intervals of $n$, showing that $\mu$ is independent of carrier concentration. Furthermore, by extrapolating the experimental dependences $\sigma(V_g)$ to zero $\sigma$, we could determine initial ($V_g=0$) concentrations of charge carriers and their type. Graphene behaved rather similarly to few-layer graphitic samples reported in [12] and is either a shallow-gap semiconductor or a small-overlap semimetal, where positive and negative gate voltages induce 2D electrons and holes, respectively, in concentrations up to ≈10$^{13}$cm$^{-2}$. Graphene exhibited typical values of $\mu$ between 2,000 and 5,000 cm$^2$/Vs. For 2D NbSe$_2$ and MoS$_2$, we measured mobilities between 0.5 and 3 cm$^2$/Vs for different samples, in agreement with mobilities for the corresponding 3D crystals at room temperature. Both 2D dichalcogenides were found to be electron conductors with $n \approx 10^{12}$ to $10^{13}$ cm$^{-2}$. Detailed studies of their conductivities as a function of temperature and $V_g$ revealed that 2D MoS$_2$ was a heavily-doped semiconductor with an activation gap of $\geq 0.6$eV while NbSe$_2$ became a semimetal. The found electron concentration in 2D NbSe$_2$ is two orders of magnitude smaller than carrier concentrations per monolayer in 3D NbSe$_2$. This indicates significant changes in the energy spectrum of NbSe$_2$ from a normal metal in 3D to a semimetal in 2D.

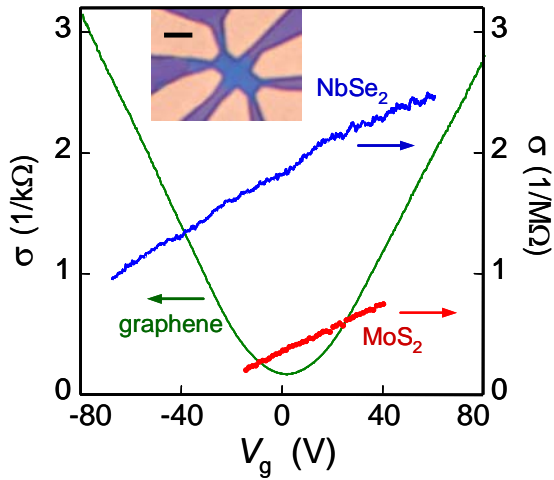

Figure 3. Electric field effect in 2D crystals. Changes in electrical conductivity σ of 2D NbSe$_2$, 2D MoS$_2$ and graphene as a function of gate voltage. The inset shows our typical devices used for such measurements: It is an optical image (in white light) of 2D NbSe$_2$ on top of an oxidized Si wafer (used as a gate electrode) with a set of Au contacts. The crystal is seen as a bluer region in the centre (scale bar – 5 μm).

In conclusion, we have demonstrated the existence of 2D atomic crystals that can be prepared by cleavage from most strongly-layered materials. It is most unexpected if not counterintuitive that isolated 2D crystals can be stable at room temperature and in air, leaving aside the fact that they maintain macroscopic continuity and such high quality that their carrier mobilities remain almost unaffected. The found class of 2D crystals offers a wide choice of new materials parameters for possible applications and promises a wealth of new phenomena usually abundant in 2D systems. We believe that, once investigated and understood, 2D crystals can also be grown in large sizes required for industrial applications, matching the progress achieved recently for the case of single-wall nanotubes [19].